\documentclass[referee]{raa}           
\usepackage{graphicx,times}
\usepackage{natbib}
\usepackage{amssymb,amsmath}
\bibpunct{(}{)}{;}{a}{}{,}

\usepackage[a4paper=true,dvipdfm=true,pagebackref=true]{hyperref}
\usepackage{color}
\hypersetup{pdftitle = The title of my PDF, pdfauthor = My name, pdfsubject= The subject, pdfkeywords = keyword1 keyword2 keyword3} 
\hypersetup{colorlinks = true, linkcolor = green, anchorcolor = red, citecolor = blue, filecolor = red, pagecolor = red, urlcolor = red}

\begin{document}

   \title{Analytical Fits to the Synchrotron Functions
}

 \volnopage{ {\bf 2012} Vol.\ {\bf X} No. {\bf XX}, 000--000}
   \setcounter{page}{1}

   \author{  M. Fouka\inst{1}, S. Ouichaoui\inst{2}   }

   \institute{ Research Center in Astronomy, Astrophysics and Geophysics, B.P. 63, Algiers observatory, Bouzar\'{e}ah, Algiers, Algeria; {\it m.fouka@craag.dz}\\
        \and
             Universit\'{e} des Sciences et de la Technologie Houari Boumedi\`{e}ne
(USTHB), Facult\'{e} de Physique, Laboratoire SNIRM, B.P. 32, El-Alia, 16111 Bab Ezzouar, Algiers, Algeria; {\it souichaoui@usthb.dz}\\
\vs \no
   {\small Received 2012 December 04; accepted 2013 XXXX 00}
}

\abstract
{
Accurate fitting formulae to the synchrotron function, $F(x)$, and its complementary function, $G(x)$, are performed and presented. The corresponding relative errors are less than $0.26\%$ and $0.035\%$ for $F(x) $ and $G(x)$, respectively. To this aim we have, first, fitted the modified Bessel functions, $K_{5/3}(x)$ and $K_{2/3}(x)$. For all the fitted functions, the general fit expression is the same, and is based on the well known asymptotic forms for low and large $x$-values for each function. It consists of multiplying each asymptotic form by a function that
tends to unity or zero for low and large $x$-values. Simple formulae are suggested in this paper, depending on adjustable parameters. The latter have been determined by adopting the Levenberg-Marquardt algorithm. The proposed formulae should be of great utility and simplicity for computing spectral powers and the degree of polarization for the synchrotron radiation, both for laboratory and astrophysical applications.   
}

   \authorrunning{M. Fouka and S. Ouichaoui }            
   \titlerunning{Analytical Fits to the Synchrotron Functions}  
   \maketitle

%
\section{Introduction}

Analytical approximate formulae are often very useful and may be
indispensable in order to avoid the computation of complicated
transcendental functions. This is the case of the modified Bessel functions
and their integrals, especially those of the second kind with fractional
order, e.g., $K_{5/3}(x)$ and $K_{2/3}(x)$, on which we focus our attention
in this contribution. We start by presenting, in section \ref{Bessel_functions}, results of fits to these two functions. Then, in section \ref{Fx_function}, we deduce the expression of the complementary synchrotron function, $G(x)=xK_{2/3}(x)$, directly from function $K_{2/3}(x)$, and report the corresponding fit to the synchrotron function, $F(x)$, before concluding in section \ref{conclusion}.

\section{Modified Bessel functions $K_{5/3}$ and $K_{2/3}$} \label{Bessel_functions}

\subsection{Definitions}

The modified Bessel functions, $I_{\pm \nu }(x)$ and $K_{\nu }(x)$, of the
first and second kind, respectively, are particular solutions of Bessel's
cylindrical differential equation, i.e., (\citealt{Abramowitz+Stegun+1965})

\begin{equation}
x^{2}\frac{d^{2}w}{dx^{2}} + x\frac{dw}{dx}-\left( x^{2}+\nu^{2} \right)w=0.
\label{eq_Bessel_diff}
\end{equation}

Function $K_{\nu }(x)$ expresses as (\citealt{Abramowitz+Stegun+1965}) 

\begin{equation}
K_{\nu }(x)=\frac{\pi }{2}\frac{I_{-\nu }(x)-I_{\nu }(x)}{\textnormal{sin}(\nu \pi )},
\label{eq_Knu}
\end{equation}

in terms of function $I_{\nu }(x)$ that writes as (\citealt{Abramowitz+Stegun+1965})

\begin{equation}
I_{\nu }(x)=\left( \frac{x}{2}\right) ^{\nu }\sum_{k=0}^{\infty }\frac{\displaystyle{\left( \frac{x}{2}\right) ^{2k}}}{k!\Gamma (\nu +k+1)},
\label{eq_Inu}
\end{equation}

in form of an ascending series involving the $\Gamma$ function. Besides,
function $K_{\nu }(x)$ can also be written as (\citealt{Abramowitz+Stegun+1965})

\begin{equation}
K_{\nu }(x)=\frac{\pi ^{\frac{1}{2}}(\frac{1}{2}z)^{\nu }}{\Gamma(\nu +\frac{1}{2})}\int_{1}^{\infty }e^{-zt}(t^{2}-1)^{\nu -\frac{1}{2}}dt,
\label{eq_Knu_int_rep}
\end{equation}

in integral representation.

Finally, this function admits the following simplified asymptotic forms (\citealt{Abramowitz+Stegun+1965})

\begin{equation}
K_{\nu }(x)\approx \left\{ 
\begin{array}{l}
\vspace{0.1cm}\displaystyle{}A_{1}(x)=\frac{1}{2}\Gamma (\nu )\left( \frac{x}{2}\right) ^{-\nu }~~~\textnormal{for}~x\ll 1 \\ 
\displaystyle{}A_{2}(x)=\sqrt{\frac{\pi }{2}}\:x^{-\frac{1}{2}}e^{-x}~~~\textnormal{for}~x\gg 1
\end{array}
\right..
\label{eq_asymp_Knu}
\end{equation}

\subsection{Fitting formulae}

In fitting a function, $f(x)$ (here, the modified Bessel functions and the
synchrotron functions), the main idea consists in expressing it in terms of
its known asymptotic forms, say $A_{1}(x)$ for low $x$-values and $A_{2}(x)$
for large $x$-values, and to put it under the form

\begin{equation}
f(x)=A_{1}(x)\delta_{1}(x)+A_{2}(x)\delta_{2}(x),
\label{eq_fit}
\end{equation}

where $\delta _{1}(x)$ and $\delta _{2}(x)$ are the functions one is looking
for, which must respectively obey the limits

\begin{equation}
\left\{ 
\begin{array}{l}
\vspace{0.1cm}\delta _{1}(x)\approx 1~~~\textnormal{for}~~~x\ll 1 \\ 
\delta _{1}(x)\approx 0~~~\textnormal{for}~~~x\gg 1
\end{array}
\right.
\label{eq_delta1_limits}
\end{equation}

and

\begin{equation}
\left\{ 
\begin{array}{l}
\vspace{0.1cm} \delta_{2}(x)\approx 0~~~\textnormal{for}~~~x\ll 1 \\ 
\delta_{2}(x)\approx 1~~~\textnormal{for}~~~x\gg 1
\end{array}
\right..
\label{eq_delta2_limits}
\end{equation}

For this purpose, we propose the following expressions:

\begin{equation}
\left\{ 
\begin{array}{l}
\vspace{0.1cm} \delta_{1}(x)= e^{H_{1}(x)} \\ 
\displaystyle{} H_{1}(x)=\sum_{k=1}^{n_{1}}a^{(1)}_{k}x^{1/k}
\end{array}
\right.
\label{eq_delta_1}
\end{equation}

and

\begin{equation}
\left\{ 
\begin{array}{l}
\vspace{0.1cm} \delta_{2}(x)= 1-e^{H_{2}(x)} \\ 
\displaystyle{} H_{2}(x)=\sum_{k=1}^{n_{2}}a^{(2)}_{k}x^{1/k}
\end{array}
\right.
\label{eq_delta_2}
\end{equation}

\begin{figure}
\centering
\includegraphics[width=10cm]{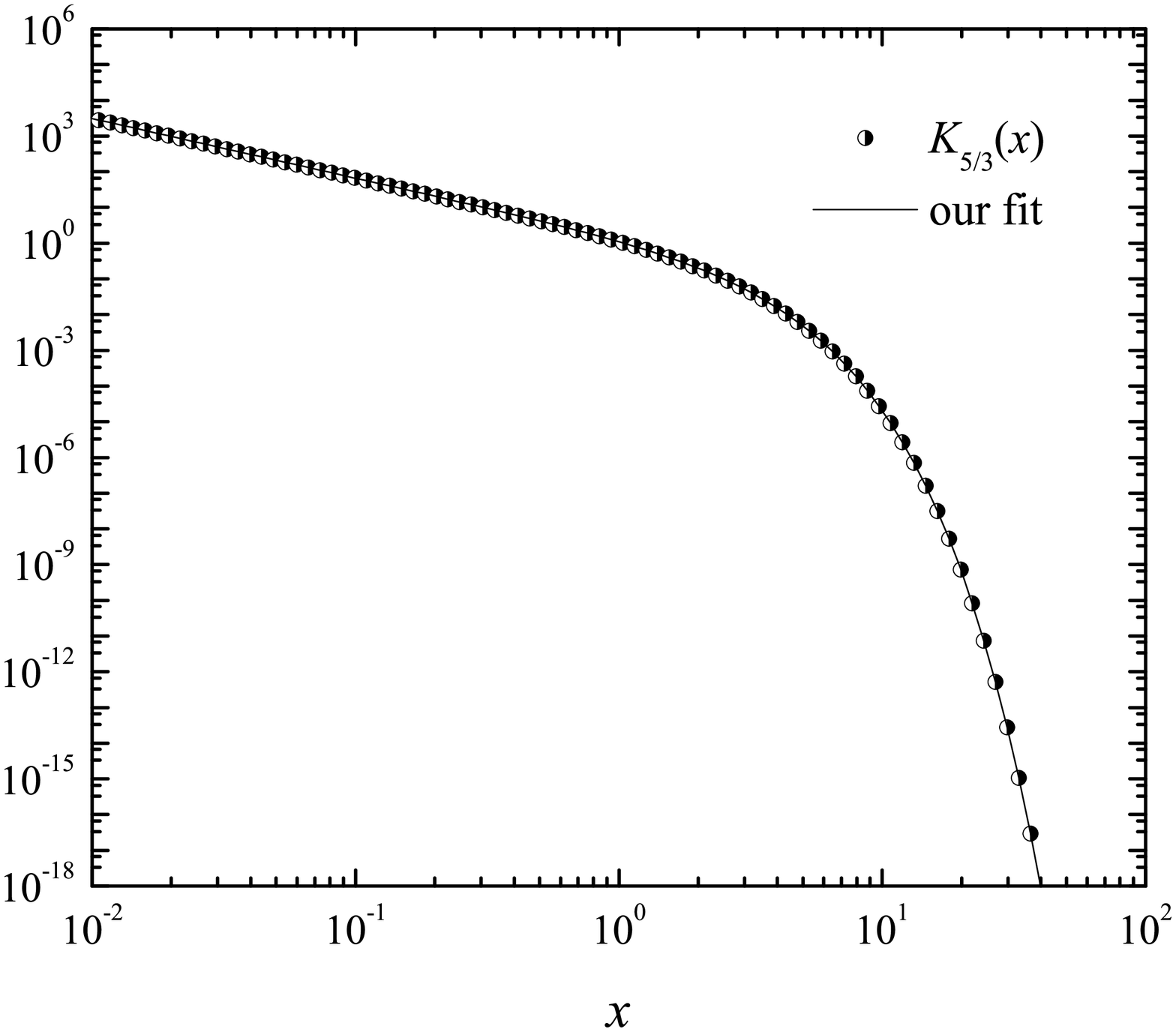}
\caption{The modified Bessel function, $K_{5/3}(x)$, together with its
corresponding fit according to equation (\ref{eq_fit}).}
\label{fig_k_5_3}
\end{figure}

\begin{figure}
\centering
\includegraphics[width=10cm]{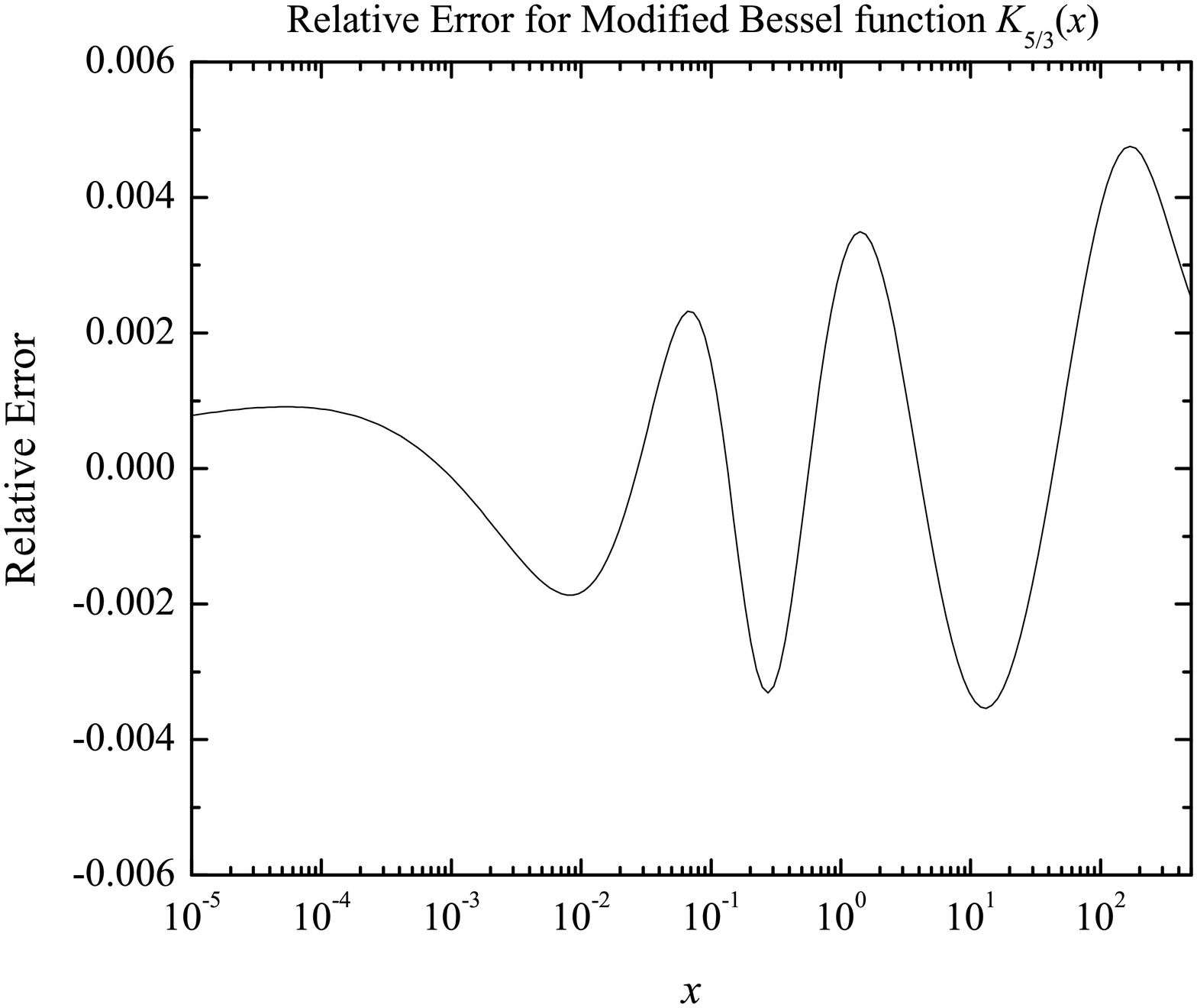}
\caption{The relative error for the modified Bessel function, $K_{5/3}(x)$,
corresponding to the set of coefficients reported by table \ref{tab_fit_K_5_3}.}
\label{fig_error_k_5_3}
\end{figure}

\begin{figure}
\centering
\includegraphics[width=10cm]{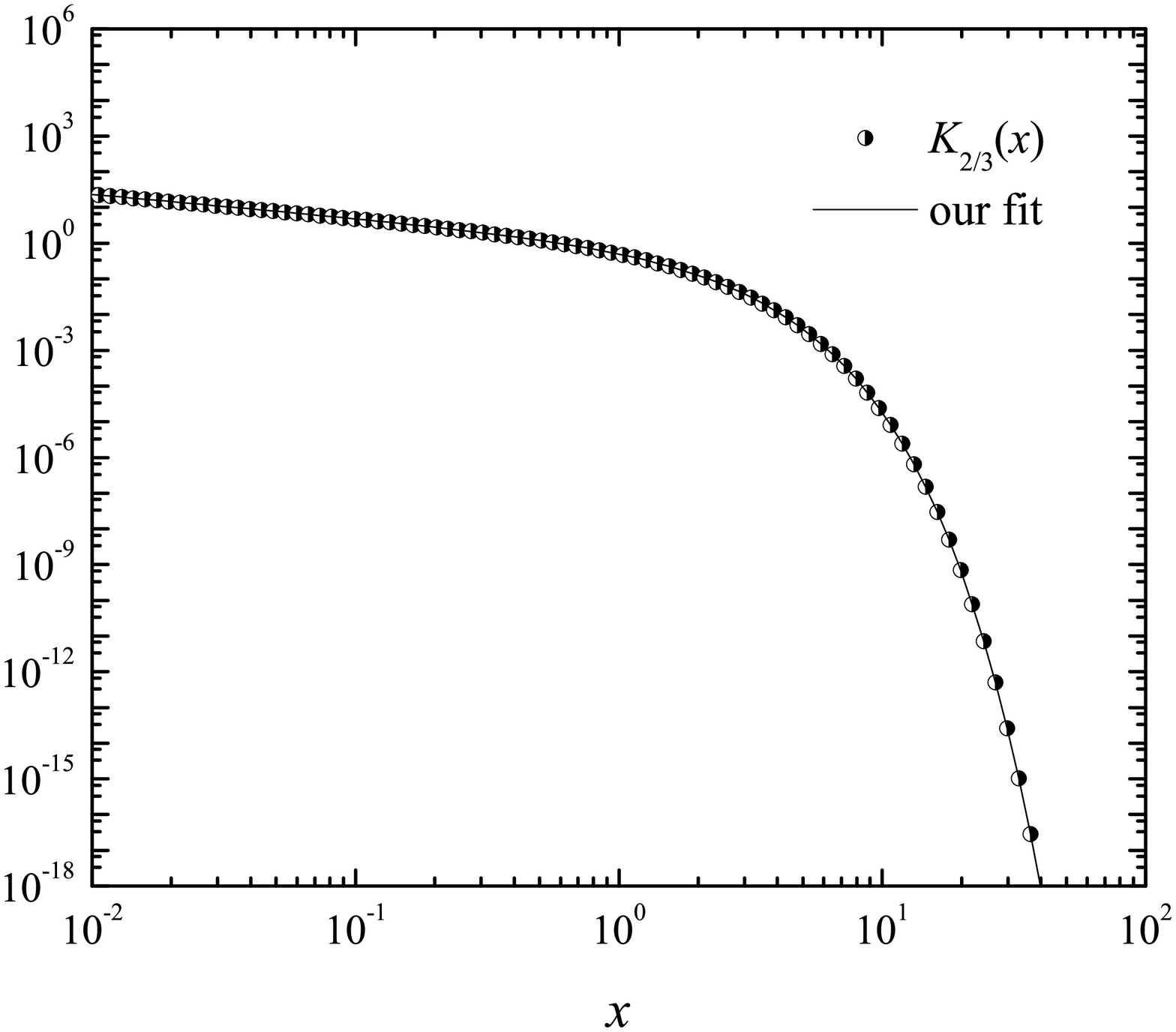}
\caption{The modified Bessel function, $K_{2/3}(x)$, together with its
corresponding fit, according to equation (\ref{eq_fit}).}
\label{fig_k_2_3}
\end{figure}

\begin{figure}
\centering
\includegraphics[width=10cm]{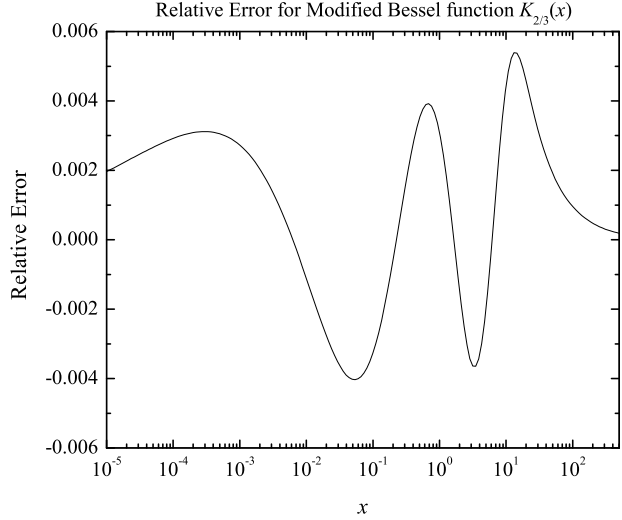}
\caption{The relative error for the modified Bessel function, $K_{2/3}(x)$,
corresponding to the set of coefficients reported by table \ref{tab_fit_K_2_3}.}
\label{fig_error_k_2_3}
\end{figure}

In order to extract coefficients $a_{k}^{(1)}$ and $a_{k}^{(2)}$ for a given
couple of orders $(n_{1},n_{2})$, we proceed by chi-squares minimization
with adopting the Levenberg-Marquardt algorithm (\citealt{Levenberg+1944}, \citealt{Marquardt+1963}), in log-log scale. The obtained fit results to functions $K_{5/3}(x)$ and $K_{2/3}(x)$ are presented in tables \ref{tab_fit_K_5_3} and \ref{tab_fit_K_2_3}, respectively, in terms of coefficients $a_{k}^{(1)}$ and $a_{k}^{(2)}$, with $n_{1}=3$ and $n_{2}=1$ and relative respective errors, $<0.48\%$ and $<0.54\%$. These fits to functions $K_{5/3}(x)$ and $K_{2/3}(x)$ are plotted in figures \ref{fig_k_5_3} and \ref{fig_k_2_3}, respectively, while the corresponding
relative errors are reported in figures \ref{fig_error_k_5_3} and \ref{fig_error_k_2_3}.

\begin{table}
\bc
\begin{minipage}[]{100mm}
\caption[]{Coefficients $a^{(1)}_{k}$ and $a^{(2)}_{k}$ for function $K_{5/3}(x)$. \label{tab_fit_K_5_3}}
\end{minipage}
\setlength{\tabcolsep}{1pt}
\small
 \begin{tabular}{ccc}
  \hline\noalign{\smallskip}
~~~$k$ ~~~   & $a^{(1)}_{k}$                         ~~~   & $a^{(2)}_{k}$         ~~~  \\
  \hline\noalign{\smallskip}
~~~$1$~~~    &  $-1.0194198041210243 $               ~~~   & $-15.761577796582387$ ~~~  \\     
~~~$2$~~~    &  $+0.28011396300530672 $               ~~~   &                       ~~~  \\
~~~$3$~~~    &  $-7.71058491739234908 \times 10^{-2}$ ~~~   &                       ~~~  \\
  \noalign{\smallskip}\hline
\end{tabular}
\tablecomments{0.86\textwidth}{ With this set of coefficients, the relative error is $<0.48 \%$.}
\ec
\end{table}

\begin{table}
\bc
\begin{minipage}[]{100mm}
\caption[]{Coefficients $a^{(1)}_{k}$ and $a^{(2)}_{k}$ for function $K_{2/3}(x)$.  \label{tab_fit_K_2_3}}
\end{minipage}
\setlength{\tabcolsep}{1pt}
\small
 \begin{tabular}{ccc}
  \hline\noalign{\smallskip}
~~~$k$~~~    & $a^{(1)}_{k}$           ~~~           & $a^{(2)}_{k}$         ~~~    \\
  \hline\noalign{\smallskip}
~~~$1$~~~    &  $-1.3746667760953621 $ ~~~           & $-0.33550751062084 $  ~~~    \\     
~~~$2$~~~    &  $+0.44040512552162292 $ ~~~          &                       ~~~    \\
~~~$3$~~~    &  $-0.15527012012316799 $ ~~~          &                       ~~~    \\
  \noalign{\smallskip}\hline
\end{tabular}
\ec
\tablecomments{0.86\textwidth}{With this set of coefficients, the relative error is $<0.54\%$.}
\end{table}

For high accuracy, we give, in table \ref{tab_fit_K_2_3_accurate}, fit
results for function $K_{2/3}(x)$, with $n_{1}=n_{2}=4$ and with a relative
error $<0.035\%$.

\begin{table}
\bc
\begin{minipage}[]{100mm}
\caption[]{Coefficients $a^{(1)}_{k}$ and $a^{(2)}_{k}$ for function $K_{2/3}(x)$. \label{tab_fit_K_2_3_accurate}}\end{minipage}
\setlength{\tabcolsep}{1pt}
\small
 \begin{tabular}{ccc}
  \hline\noalign{\smallskip}
~~~$k$~~~    & $a^{(1)}_{k}$  & $a^{(2)}_{k}$\\
  \hline\noalign{\smallskip}
~~~$1$~~~    &  $-1.0010216415582440$~~~~~~~~~      & $-0.2493940736333195 $ ~~~  \\     
~~~$2$~~~    &  $+0.88350305221249859$~~~~~~~~~     & $+0.9122693061687756 $ ~~~  \\
~~~$3$~~~    &  $-3.6240174463901829$~~~~~~~~~      & $+1.2051408667145216 $ ~~~  \\
~~~$4$~~~    &  $+0.57393980442916881$~~~~~~~~~     & $-5.5227048291651126 $ ~~~  \\
  \noalign{\smallskip}\hline
\end{tabular}
\ec
\tablecomments{0.86\textwidth}{With this set of coefficients, the relative error is $<0.035\%$.}
\end{table}

\section{Synchrotron Functions}\label{Fx_function} 
\subsection{Definitions}
The synchrotron functions, $F(x)$ and $G(x)$, are
defined by (\citealt{Westfold+1959, Jackson+1962, Rybicki+Lightman+1979, Fouka+Ouichaoui+2009}):

\begin{equation}
\left\{ 
\begin{array}{l}
\vspace{0.1cm}
\displaystyle{}F(x)=x\int^{\infty}_{x}K_{5/3}(x^{\prime})dx^{\prime} \\ 
G(x)=xK_{2/3}(x) 
\end{array}
\right..
\label{eq1}
\end{equation}

Function $G(x)$ is called the complementary synchrotron function and is
sometimes noted $F_{p}(x)$ (\citealt{Westfold+1959}). The corresponding simplest
asymptotic forms of these functions have the following expressions (\citealt{Westfold+1959, Rybicki+Lightman+1979}):

\begin{equation}
F(x)\approx \left\{ 
\begin{array}{l}
\vspace{0.1cm}
\displaystyle{}F_{1}x^{\:1/3}~~~\textnormal{for}~~~x\ll 1 \\ 
\displaystyle{}F_{2}e^{-x}x^{1/2}~~~\textnormal{for}~~~x\gg 1
\end{array}
\right.
\label{eq_F_asympt}
\end{equation}

and

\begin{equation}
G(x) \approx \left\{ 
\begin{array}{l}
\vspace{0.1cm}
\displaystyle{}G_{1} x^{\:1/3}~~~\textnormal{for}~~~ x\ll 1  \\ 
\displaystyle{}G_{2} e^{-x}x^{1/2}~~~\textnormal{for}~~~x\gg 1
\end{array}
\right.,
\label{eq_G_asymp}
\end{equation}

where $F_{1}=\pi 2^{5/3}/\sqrt{3}\Gamma (1/3)$, $F_{2}=\sqrt{\pi /2}$ and $%
G_{1}=F_{1}/2$ , $G_{2}=F_{2}$.

\subsection{Fitting formulae}
\begin{figure}
\centering
\includegraphics[width=10cm]{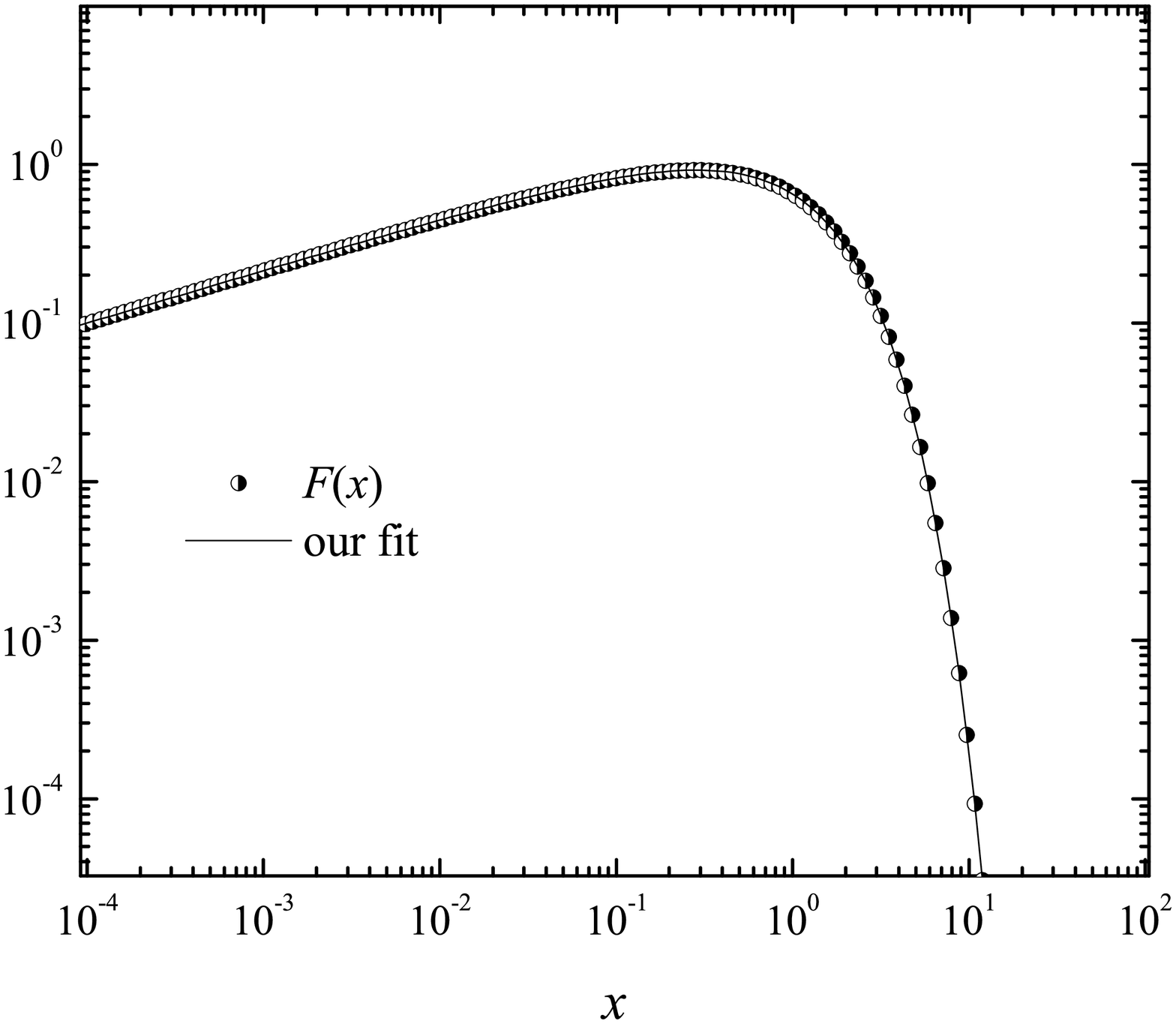}
\caption{The synchrotron function, $F(x)$, together with its corresponding
fit according to equation (\protect\ref{eq_fit}).}
\label{fig_Fx}
\end{figure}

\begin{figure}
\centering
\includegraphics[width=10cm]{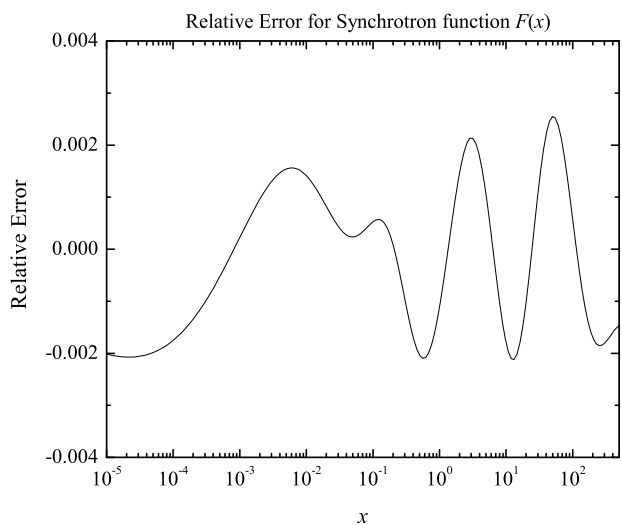}
\caption{The relative error for the synchrotron function, $F(x)$, corresponding to the set of coefficients reported by table \protect\ref{tab_fit_Fx}.}
\label{fig_error_Fx}
\end{figure}

Function $G(x)$ can be easily derived directly from the fit to function $%
K_{2/3}(x)$. One has just to multiply the latter by variable $x$. For
fitting function $F(x)$ , we proceed in the same way as for the modified
Bessel functions, i.e., putting it under the form given by equation (\ref%
{eq_fit}). We have just to consider the corresponding asymptotic forms given
by equation (\ref{eq_F_asympt}). The corresponding fit coefficients are
reported in table \ref{tab_fit_Fx}. With these coefficients, the relative
error is $<0.26\%$. Function $F(x)$ is plotted in figure \ref{fig_Fx},
together with the corresponding fit while the relative error is reported in
figure \ref{fig_error_Fx}, as a function of variable $x$.

\begin{table}
\bc
\begin{minipage}[]{100mm}
\caption[]{Coefficients $a^{(1)}_{k}$ and $a^{(2)}_{k}$ for the synchrotron function $F(x)$. \label{tab_fit_Fx}}
\end{minipage}
\setlength{\tabcolsep}{1pt}
\small
 \begin{tabular}{ccc}
  \hline\noalign{\smallskip}
~~~$k$~~~~~~    &  $a^{(1)}_{k}$            ~~~~~~~~~   & $a^{(2)}_{k}$                         ~~~  \\
  \hline\noalign{\smallskip}
~~~$1$~~~~~~    &  $-0.97947838884478688  $ ~~~~~~~~~   & $-4.69247165562628882\times 10^{-2} $ ~~~  \\     
~~~$2$~~~~~~    &  $-0.83333239129525072  $ ~~~~~~~~~   & $-0.70055018056462881 $               ~~~  \\
~~~$3$~~~~~~    &  $+0.15541796026816246  $ ~~~~~~~~~   & $1.03876297841949544\times 10^{-2}  $ ~~~  \\
  \noalign{\smallskip}\hline
\end{tabular}
\ec
\tablecomments{0.86\textwidth}{ With this set of coefficients, the relative error is  $<0.26\%$.}

\end{table}

\section{Conclusion}\label{conclusion} 
We have presented analytical fit formulae with good accuracies for the synchrotron function, $F(x)$, and its complementary function, $G(x)$, based on their known asymptotic forms for low and large $x$-values. We propose these formulae for the aim of directly and simply computing these transcendental functions with avoiding fastidious
calculations. The derived general fit formulae can thus be used to evaluate the modified Bessel functions of any order: integer or non integer. Finally, these fit formulae should be of great help for computing quantities of interest to synchrotron radiation such as, e.g., the spectral power and the degree of polarization, both for laboratory and astrophysical applications.

\normalem
\begin{acknowledgements}
Thanks to Dr. Y. Damerdji for kind help on the use of the Levenberg-Marquardt method. This work was supported by the {\it National Administration of Scientific Research NASR-DZ, of Algeria}, in the framework of {\it National Projects of Research (NPR)}.
\end{acknowledgements}

\bibliographystyle{raa}
\bibliography{bibtex}

\end{document}